# Heliosheath Properties Measured from a *Voyager* 2 to *Voyager* 1 Transient

J. S. Rankin[1], D. J. McComas[1], J. D. Richardson[2], N. A. Schwadron[3]
[1]Department of Astrophysical Sciences, Princeton University, Princeton, NJ, USA
[2]Kavli Center for Astrophysics and Space Science, Massachusetts Institute of Technology, Cambridge, MA
[3]University of New Hampshire, Durham, NH, USA

## Abstract

In mid-2012, a GMIR observed by *Voyager 2* crossed through the heliosheath and collided with the heliopause, generating a pressure pulse that propagated into the very local interstellar medium. The effects of the transmitted wave were seen by *Voyager 1* just 93 days after its own heliopause crossing. The passage of the transient was accompanied by long-lasting decreases in galactic cosmic ray intensities that occurred from ~2012.55 to ~2013.35 and ~2012.91 to ~2013.70 at *Voyager 2* and *Voyager 1*, respectively. Omnidirectional ($\gtrsim$ 20 MeV) proton-dominated measurements from each spacecraft's Cosmic Ray Subsystem reveal a remarkable similarity between these causally-related events, with a correlation coefficient of 91.2% and a time-lag of 130 days. Knowing the locations of the two spacecraft, we use the observed time-delay to calculate the GMIR's average speed through the heliosheath (inside the heliopause) as a function of temperature in the very local interstellar medium. This, combined with particle, field, and plasma observations enables us to infer previously unmeasured properties of the heliosheath, including a range of sound speeds and total effective pressures. For a nominal temperature of ~20,000 K just outside the heliopause, we find a sound speed of $314 \pm 32$ km/s and total effective pressure of $267 \pm 55$ fPa inside the heliopause. We compare these results with the *Interstellar Boundary Explorer's* data-driven models of heliosheath pressures derived from energetic neutral atom fluxes (the globally distributed flux) and present them as additional evidence that the heliosheath's dynamics are driven by suprathermal energetic processes.

## 1. Introduction

The sun moves with respect to the Local Interstellar Medium (LISM) at a speed of ~26 km/s. Its supersonically-expanding solar wind inflates a bubble, known as the heliosphere, in the surrounding interstellar material. At the Termination Shock (TS), the solar wind's speed abruptly slows from supersonic (~400 km/s) to subsonic (~100 km/s; Parker, 1961). Beyond the TS is the heliosheath, a turbulent region characterized by low plasma densities (~0.002 cm$^{-3}$; Richardson et al. 2008; Richardson & Decker 2015) and weak magnetic fields (~0.1 nT; Burlaga et al. 2013, 2018). Eventually, these subsonic flows get deflected, and by the heliopause (HP), the heliospheric plasma, field, and energetic particles ultimately must achieve a pressure balance with the local interstellar medium (Davis 1955; Parker 1963).

In late 2004, *Voyager 1* became the first spacecraft to cross the TS, at a radial distance of 94 AU (Stone et al. 2005; Decker et al. 2005; Burlaga et al. 2005; Gurnett & Kurth 2005). One surprising discovery was that the Anomalous Cosmic Ray spectrum was still modulated at the shock, which was quickly explained by McComas & Schwadron (2006) by its blunt shape and acceleration back along its flanks and tail, with subsequent support by Kóta & Jokipii (2008) and Schwadron et al. (2008). Alternative explanations for ACR acceleration have also been proposed, including heliosheath mechanisms such as magnetic reconnection (e.g. Drake et al. 2010; Zank et al. 2015), compressive turbulence (Strauss et al. 2010) and pump acceleration (Fisk & Gloeckler 2014). *Voyager 1*'s TS encounter was followed in 2007 by the *Voyager 2*, which, at 84 AU, revealed that the TS was not just blunt, but also asymmetric (Stone et al. 2008; Decker et al. 2008; Burlaga et al. 2008; Gurnett & Kurth 2008; Richardson et al. 2008). This second TS crossing was marked by another surprising discovery: the TS was much weaker than expected, and the heliosheath was much colder than some models predicted (~180,000 K vs. ~1,000,000 K; Zank et al. 1996; Richardson et al. 2008; Richardson 2008). Data from *Voyager 2*'s Plasma Science (PLS) instrument revealed that over 80% of the solar wind's flow energy was being transferred to a higher-energy part of the plasma distribution, beyond PLS's thermal measurement range (>6 keV; Bridge et al. 1977; Richardson et al. 2008). Since then, both *Voyagers* have crossed the HP (*Voyager 1* in 2012, at 122 AU; *Voyager 2* in 2018, at 119 AU) and have continued outward into the Very Local Interstellar Medium[1] (VLISM; Stone et al. 2013; Krimigis et al. 2013; Webber & McDonald 2013; Burlaga et al. 2013; Gurnett et al. 2013; Brown et al. 2018).

Meanwhile, in 2008, the *Interstellar Boundary Explorer* (*IBEX*) began its mission to characterize the global properties of the heliosphere and its interaction with the interstellar medium (McComas et al. 2009a, 2009b). *IBEX* makes remote plasma observations by imaging neutral particles that are generated via charge-exchange between interstellar neutrals and energetic solar wind and pickup ions. The resulting energetic neutral atoms (ENAs) travel back into the interplanetary medium and are directly measured by *IBEX* at 1 AU. Among its first results, *IBEX* discovered a "Ribbon" of enhanced ENA emissions organized by the LISM magnetic field (see e.g., McComas et al. 2017a and reference therein for details of *IBEX* discoveries).

The suprathermal distribution of energetic particles is dominated by pickup ions (PUIs; e.g. Vasyliunas & Siscoe, 1976; Isenberg et al. 1986), which are formed by incoming interstellar neutral atoms that are ionized and carried out with the solar wind. Using observations from the Solar Wind Around Pluto (SWAP) instrument (McComas et al. 2008a) on *New Horizons*, McComas et al. (2017b) reported the properties of PUIs from 20 to 38 AU and showed that by ~20 AU, PUIs provide the dominant internal pressure in the solar wind. Extrapolating values out to the TS, McComas et al. (2017b) also inferred a ratio of PUI pressure to SW dynamic pressure of ~0.16, a value ~2.5 times larger than expected.

It is the PUIs that are responsible for weakening the termination shock and causing the solar wind flow to become subsonic (Zank et al. 1996, 2010; Richardson et al. 2008; Prested et al. 2008; Wu et al. 2009; Mostafavi et al. 2018) – despite their contributing to only about ~20% of the total mass density of the plasma (e.g., Richardson et al. 2008; Desai et al. 2014; McComas et al. 2017b). IBEX observations reveal that the plasma environments of the heliosphere and VLISM are strongly coupled through charge exchange processes (see, e.g. Zank et al. 2016; McComas et al. 2017a) and demonstrate that PUIs carry a significant faction of the heliosheath's total pressure. To date, only the core thermal component of the subsonic SW plasma has been directly measured in the heliosheath. Therefore, obtaining a better understanding of the heliosheath plasma parameters such as sound speed and total effective pressure is critical for understanding not only the unmeasured distributions in the heliosheath, but also the global properties, pressure balance, and overall interaction of the heliosphere and interstellar medium.

---

[1] We adopt the convention of Zank (2015) and Zank et al. (2017) and use an updated definition of "VLISM", which they propose as: "that region of the interstellar medium surrounding the Sun that is modified or mediated by heliospheric processes or material".

Returning to the inner solar system, coronal mass ejections (CMEs), co-rotating interaction regions (CIRs), and other transients produced by the sun's activity can coalesce to form large merged interaction regions (MIRs) and even larger global merged interaction regions (GMIRs) (e.g. Burlaga 1995). Passages of these large-scale structures are often marked by intense magnetic fields, enhanced pressures, and increases of low-energy particles that are accelerated by the compressed fields. The short-lived particle enhancements are typically followed by longer-duration decreases of energetic particle intensities, including galactic cosmic rays (GCRs). GMIRs are capable of persisting beyond the termination shock, propagating through the heliosheath, and generating compressive waves at the HP boundary which continue outward into the VLISM (e.g., Gurnett et al. 1993, Whang & Burlaga 1994). Such events have been directly observed in the heliosheath by *Voyager 2* (Burlaga & Ness 2016; Richardson et al. 2017; Burlaga et al. 2018a, 2019) and their transmission across the HP has been modeled extensively via time-dependent simulations that are often informed by solar wind observations (Zank & Müller 2003; Washimi et al. 2011, 2017; Liu et al. 2014; Fermo et al. 2015; Kim et al. 2017).

*Voyager 1* has observed multiple signatures of transient disturbances in the VLISM on all four of its working instruments. The Plasma Wave (PWS) instrument detected several locally-generated electron plasma emissions, typically preceded by GCR intensity enhancements measured by the Cosmic Ray Subsystem (CRS) and Low Energy Charged Particle Instruments (LECP; Gurnett et al. 2013, 2015). Following these precursors, the magnetometer (MAG) reported several disturbances in the form of weak, quasi-perpendicular, laminar changes to the magnetic field (Burlaga et al. 2013; Burlaga & Ness 2016; Burlaga et al. 2019). LECP and CRS additionally observed an unexpected anisotropy in the GCRs characterized by a time-varying depletion of particles with near 90° pitch angles (Krimigis et al. 2013; Rankin et al. 2019). There is a suggestion that the transient disturbances and GCR anisotropies are related (e.g. Jokipii & Kóta 2014; Kóta & Jokipii 2017). Nonetheless, many aspects of the GCR anisotropies are not yet fully understood.

In this study, we investigate intensity changes of GCRs in the heliosheath and VLISM, using omnidirectional rates ($\gtrsim$ 20 MeV; proton-dominated) measured by CRS on *Voyager 2* and *Voyager 1*, respectively. These observations enable us to infer otherwise unmeasured plasma parameters in the heliosheath including the sound speed and total effective pressure. In Section 2, we examine the GCR response to a GMIR that likely passed through the heliosheath and collided with the HP, generating a pressure pulse that was transmitted into the VLISM. In Section 3, we use the GMIR's derived average speed in combination with particle, field, and plasma observations to estimate a range of average sound speeds and total effective pressures in the heliosheath, including that of the pickup ions. In Section 4, we compare our heliosheath pressures to current observations and models, and in Section 5, we discuss our findings in light of recent *Voyager* and *IBEX* observations.

# 2. Observations of GCR Disturbances in the Heliosheath and VLISM

## 2.1 Causally-Related GCR Disturbances at *Voyager 1* and *Voyager 2*

We perform a cross-correlation analysis to compare the time evolution of *Voyager 1*'s first large GCR anisotropy episode in the VLISM (~2012.91 to ~2013.70; Rankin et al. 2019) to *Voyager 2's* measurements in the heliosheath from 2010.0 to 2013.7 (see Appendix A for details). Our analysis identifies a significantly-correlated GCR disturbance at *Voyager 2* that has a corresponding time window of ~2012.55 to ~2013.35, a correlation coefficient of $CC = 91.2\% \pm 2.1\%$ and a time-lag of $\Delta t = 130 \pm 5$ days. The intensity profiles for these events are shown in Figure 1. Their remarkable similarity with a delay indicative of a reasonable propagation time indicates that they are almost certainly causally related. Our finding that the structure of the GCR disturbances is preserved in response to the pressure wave is not unexpected since they are responding to similar patterns of compression, despite the very different plasma environments of the heliosheath and VLISM. The wave inside the heliosheath modulates the heliopause much like a speaker membrane and launches the wave into the medium beyond in such a way that it preserves the waveform's structure. To use another physical analogy, a non-dispersive disturbance (such as a fast magnetosonic wave) has equal phase and group velocities, which enables it to propagate without deformation. Since the wave's speed in the heliosheath is undoubtedly much faster than in the VLISM, structural similarities could imply that a pressure pulse elongates through its interaction with the HP. Moreover, once it arrives in the VLISM, a pressure pulse that is initially subsonic may eventually steepen to form a shock. Given that the event in the current study was detected within 1 AU of *Voyager 1*'s HP crossing, a follow-on investigation using later events is needed to acquire a more-definitive understanding of these details. Finally, waves and shocks likely differ in their transmittance across the HP, as demonstrated by Washimi et al. (2011); therefore, the nature of the GCR disturbances might also depend on whether the heliosheath event originated from a wave or a shock.

In principal, changes to GCR intensity in the heliosheath can occur due to local phenomena (e.g. a propagating GMIR), or due to global variations (e.g. solar modulation). While the former might be accompanied by enhancements in the magnetic field and plasma, the latter might only affect GCRs as opposed to ACRs that come from magnetic connection to the flanks and tail of the TS (McComas & Schwadron 2006; Schwadron et al. 2008; Kóta & Jokipii 2008; Cummings et al. 2013; Stone et al. 2017). This particular GCR disturbance at *Voyager 2* was caused by a wave and appears to be local; ACRs exhibit a similar behavior, and the energetic particle response occurs near the time-vicinity MIR "C" identified by Richardson et al. (2017). Meanwhile in the VLISM, the onset of *Voyager 1*'s late-2012 anisotropy event (DOY 331 of 2012) coincided with what is described as a thick, laminar, collisional, subcritical shock observed by MAG on DOY 335 of 2012 (~2012.92; Burlaga et al. 2013; Burlaga & Ness 2016; Mostafavi & Zank 2018; Rankin et al. 2019) and was preceded by roughly month-long electron plasma oscillations (Gurnett et al. 2013, 2015). Our observations therefore support Richardson et al. (2017)'s hypothesis – this particular GMIR observed by *Voyager 2* likely crossed through the heliosheath, collided with the HP, and transmitted a pressure pulse that subsequently drove the interstellar transients observed at *Voyager 1*.

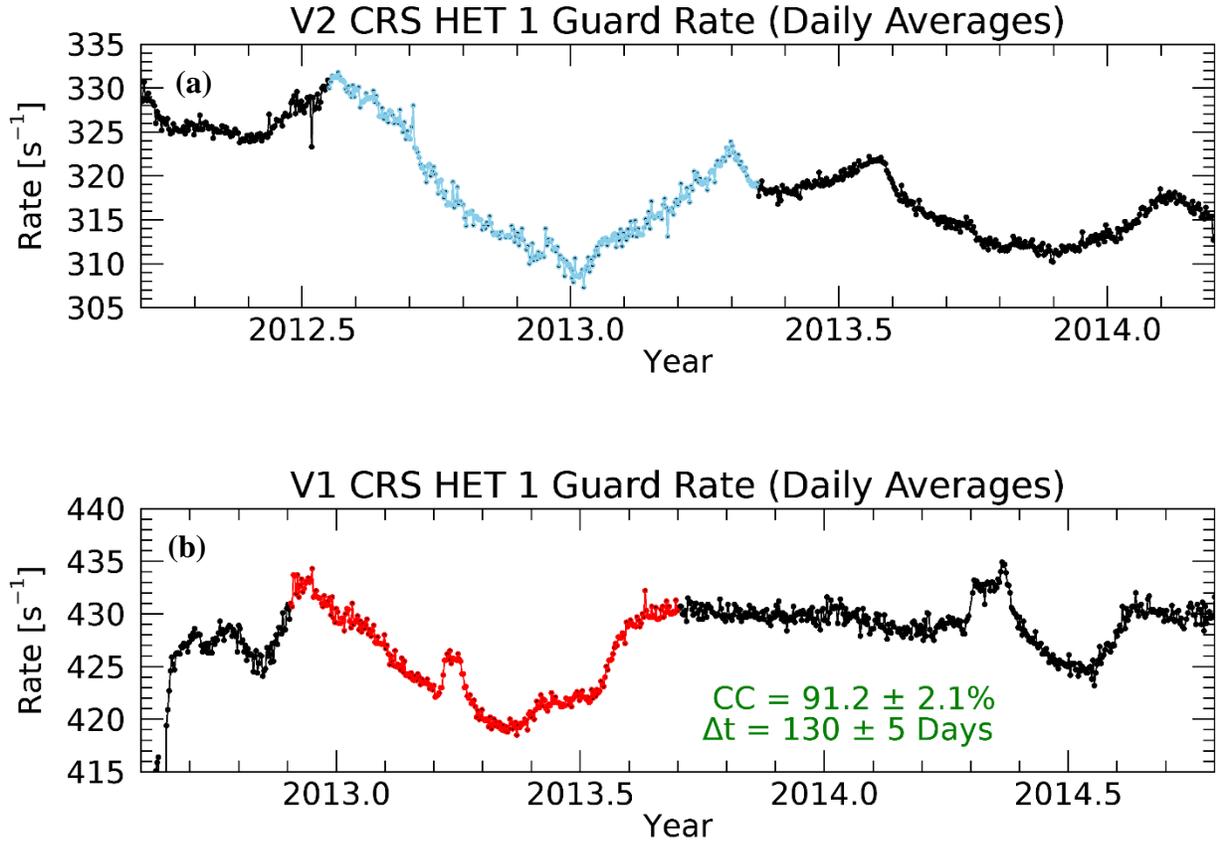

**Figure 1.** A comparison of *Voyager 2* (a) and *Voyager 1* (b) CRS GCR observations taken in the heliosheath and VLISM, respectively. The GMIR that caused the GCR intensity decrease at *Voyager 2* (blue) likely passed through the heliosheath and collided with the HP, transmitting a pressure pulse that drove the event observed by *Voyager 1* (red). Data shown are from CRS's omnidirectional guard rates measured by the HET 1 telescopes (≳ 20 MeV; proton-dominated); for more information about CRS, see Stone et al. (1977).

## 2.2 Propagation Speed of the GMIR through the Heliosheath

Knowing the coordinates of each spacecraft, the time delay between the two GCR disturbances, and the approximate location of the HP, we derive a relationship for the GMIR's radial propagation speed through the heliosheath:

$$v_{HS} = \left[\Delta t - \frac{\Delta R_{VLISM}}{v_{VLISM}}\right]^{-1} \Delta R_{HS} \quad (1),$$

where $\Delta t = 130 \pm 5$ days, $v_{VLISM}$ is the transmitted wave's propagation speed through the VLISM, and $\Delta R_{HS}$ and $\Delta R_{VLISM}$ represent the radial distances that the wave travelled through the heliosheath and VLISM – 22.53 AU and 0.91 AU, respectively. These distances were obtained using a HP location of 121.58 AU, which was measured at the time of Voyager 1's crossing on DOY 238 of 2012.

The onset of the GCR disturbance at *Voyager 1* occurred only 93 days after the spacecraft encountered the HP; however, the HP is not static. Models indicate that it could move several AU during an 11-year solar cycle (Washimi et al. 2011; Zank 2015; Pogorelov et al. 2017), as the result of transients (Pogorelov et al. 2012a; Washimi et al 2017) or as a result of sudden increases (or decreases) of solar wind dynamic pressure, such as that recently observed by *IBEX* (McComas et al. 2018; Zirnstein et al. 2018). If we assume an average HP motion of ±1 AU/yr (Washimi et al. 2017), we estimate an uncertainty of ±0.25 AU over this shorter time period, which we include in our analysis.

The fastest magnetohydrodynamic (MHD) wave propagation speed in the VLISM, $v_{VLISM}$, is a function of both magnetosonic speed and the radial bulk flow speed of the plasma and is mostly perpendicular to the field (e.g. Burlaga & Ness 2016; Zank et al. 2017). Therefore:

$$v_{VLISM} \cong v_{bulk1} + \sqrt{v_{A1}^2 + cs_{VLISM}^2} \quad (2).$$

Since *Voyager 1* does not have a working plasma instrument, the plasma flow speeds in this region have not been directly measured. However, since the spacecraft's longitude is within a few degrees of the nose (Lallement et al. 2005; McComas et al. 2015) and the disturbance was measured close to the HP, the radial bulk flow speed is likely very small, if not negligible (e.g. Holzer 1989; Pogorelov et al. 2012b). For now, we proceed assuming $v_{bulk1} = 0$; we will include alternative values with our final results in Section 3.

The Alfvén speed in the VLISM is informed by *Voyager 1's* measurements of 1) the average field strength measured by MAG: |B| = 0.48 ± 0.03 nT (Burlaga & Ness 2016) and 2) the plasma density acquired from PWS's plasma oscillations during the relevant time period: $\rho = 0.07 \pm 0.01$ cm$^{-3}$ (Gurnett et al. 2015). From these values, we find: $v_A = 39.6 \pm 3.8$ km/s. The sound speed is less certain, since *Voyager 1* cannot directly measure temperature. Nonetheless, we assume $\gamma = 5/3$ and use temperatures ranging from 7,500 K (that of the pristine LISM; McComas et al. 2015), to 40,000 K, which is about the highest value that models are able to obtain near the HP (Zank et al. 2013; Izmodenov et al. 2014; Heerikhuisen et al. 2016; Pogorelov et al. 2017).

Incorporating the above-described effects and using the approximation that *Voyager 1* and *Voyager 2* view the same radially-expanding pressure front, we find that the speed of the pressure pulse through the VLISM ranges from $v_{VLISM} = 42.1 \pm 4.0$ to $51.6 \pm 4.9$ km/s, and the speed of the GMIR through the heliosheath ranges from $v_{HS} = 421 \pm 43$ to $392 \pm 40$ km/s for VLISM temperatures of 7,500 and 40,000 K, respectively.

## 3. Heliosheath Sound Speed and Total Effective Pressure

The GMIR's speed through the heliosheath reflects a combination of the plasma's bulk flow and magnetosonic speeds, $v_{HS} = v_{bulk\,2} + v_{MS}$. Since *Voyager 2* has a working plasma instrument and therefore the Alfvén speed ($v_{A2}$) and radial flow speed ($v_{bulk2}$) are provided by measurements, we can, for the first time, establish observational limits for an average sound speed in the heliosheath, $cs_{HS}$:

$$cs_{HS} \cong \sqrt{(v_{HS} - v_{bulk2})^2 - v_{A2}^2} \quad (3).$$

Our results yield a range of values from $cs_{HS} = 329 \pm 34$ to $299 \pm 31$ km/s, derived from plasma parameters averaged over the distance covered by the GMIR (from ~99 to ~121 AU) and again dependent upon VLISM temperatures.

Moreover, the sound speed is a function of the total effective pressure and density in the heliosheath inside the heliopause:

$$P_{total} = \frac{cs_{HS}^2(\rho_{total})}{\gamma} \quad (4).$$

As one might expect, the plasma density in this region is dominated by thermal ions, with an important additional component of PUIs ($\rho_{PUI} \cong 0.20\, \rho_{total}$; e.g., Richardson et al. 2008; Desai et al. 2014; McComas et al. 2017). In contrast, the total pressure is dominated by PUIs, with some small contribution from thermal ions (e.g., Richardson et al. 2008; Zank et al. 2010; McComas et al. 2017) and ACR's (e.g., Decker et al. 2015; Guo et al. 2018). Letting $\rho_{total} \cong \rho_{thermal}/0.8$ and $\gamma = 5/3$, we calculate an average total pressure that falls between $P_{total} = 292 \pm 60$ and $242 \pm 50$ fPa.

Figure 2 shows both quantities plotted over the full range of VLISM temperatures, with outer bands indicating systematic offsets for non-zero radial bulk flows in the VLISM. The full results of our calculations are listed in Table 2 of Appendix B, with relevant observational parameters included in Tables 3 and 4.

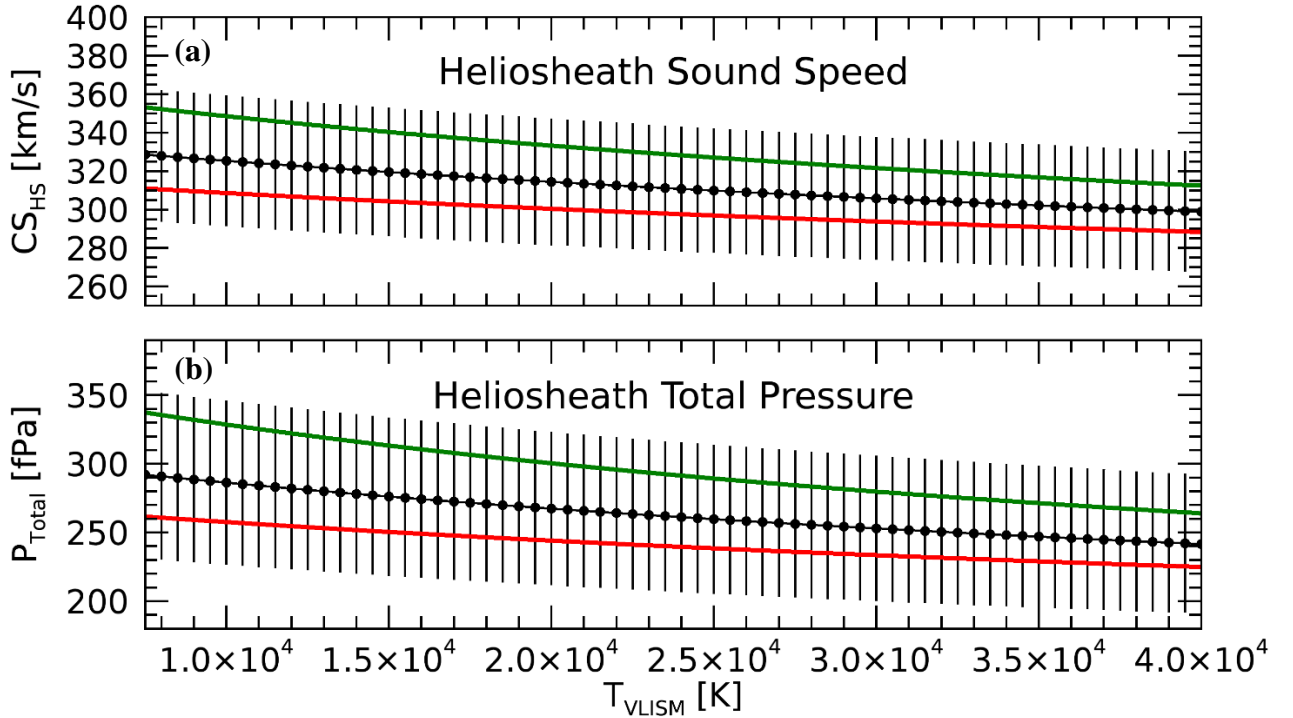

**Figure 2.** Derived sound speeds (a) and total pressures (b) in the heliosheath as a function of VLISM temperatures ranging from 7,500 K (that of the unperturbed LISM) to 40,000 K. The GCR intensity depletion associated with the transmitted wave in the VLISM occurred at *Voyager 1* when it was within 1 AU beyond its HP crossing. Our main calculation (black) assumes zero radial bulk flow speed in the VLISM, while the colored bands show the systematic offsets for radial bulk flow speeds of -5 km/s (green) and +5 km/s (red).

## 4. Comparison of Heliosheath Pressures with Current Models and Observations

In Table 1, we list our results along with heliosheath pressures derived from data-driven models. We report a nominal value of $P_{Total} = 267 \pm 55$ fPa for a temperature of $T_{VLISM} = 20{,}000$ K (Table 2, Appendix B; e.g., Zank et al. 2010). During the first year of *IBEX* observations, Funsten et al. (2009) developed a method for converting measurements of heliospheric neutral fluxes to line-of-site (LOS) integrated pressures. Schwadron et al. (2011) further improved these techniques and applied them to *IBEX*'s full-sky maps of the globally-distributed flux – the component of the flux controlled by the heliosheath's properties and processes of the of full-sky ENA emission data. Their study was extended (Schwadron et al. 2014) to include the first five years of *IBEX* observations. In conjunction with LOS-integrated pressure maps averaged over this longer time-frame, these authors also calculated the LOS-integrated pressure for a time period that was consistent with *Voyager 1*'s late-2012 HP crossing (correcting for solar-wind transport time). In doing so, they found that the pressure towards the nose had decreased significantly (by ~19%), likely as the result of the heliosheath's latent response to a solar wind pressure decrease observed at 1 AU from 2006 to 2010 (e.g., McComas et al. 2008b; Richardson & Wang 2011). Our observations may reflect this pressure decrease, owing to their temporal proximity to the affected time-frame. Therefore, in Table 1, we also consider the case for which our total pressure is scaled by a factor of 1.2. Finally, we note that Zirnstein et al. (2018)'s simulation is that of the heliosheath's response to a 50% increase in solar wind dynamic pressure in late 2014 that likely arrived in the heliosheath by early to late 2016 (McComas et al. 2018); this pressure increase occurs as a future event with respect to our current work but serves as an additional example of non-negligible temporal variability.

| Reference | $P_{Magnetic}$ | $P_{Thermal}$ | $P_{PUI}$ | $P_{SW\ Dynamic}$ | $P_{ACRs/GCRs}$ | $P_{Total}$ |
|---|---|---|---|---|---|---|
| Current Study (20,000 K) | 5.35* fPa (*V2* PLS) | 3.12* fPa (*V2* PLS) | ~160 fPa (~0.6 × $P_{Total}$) | 28.1* fPa (*V2* PLS) | ~81 fPa (~0.3 × $P_{Total}$) | ~270 fPa |
| 1.2 × Current Study (20,000 K) | --- | --- | ~190 fPa (~0.6 × $P_{Total}$) | --- | --- | ~320 fPa |
| Livadiotis et al. 2013 | --- | (0.06 × $P_{Total}$) | (0.63 × $P_{Total}$) | --- | (0.31 × $P_{Total}$) | ~210 fPa |
| Fuselier et al. 2012 | --- | --- | ~110 fPa (*Voyager* pixels; ~0.8 to 4 eV) | --- | --- | --- |
| Schwadron et al. 2011 | --- | --- | ~130 fPa (*Voyager 1*) ~110 fPa (*Voyager 2*) | --- | --- | ~190** fPa |
| Schwadron et al. 2014 | --- | --- | ~110 fPa (5-year; Fig. 9) ~80 fPa (2013; Fig. 18) | --- | --- | --- |
| Zirnstein et al. 2017 | --- | --- | ~100*** fPa | --- | --- | --- |
| Guo et al. 2018 | --- | --- | --- | --- | ~52 fPa (hypothetical energetic particle; 5 KeV to 242 MeV) | -- |
| Zirnstein et al. 2018 | --- | --- | ~175**** fPa | --- | --- | --- |

**Table 1.** A summary of pressures obtained from derived and observed (*) parameters in the heliosheath for this current work compared to data-driven models (dashed lines indicate no data). **From a 1D mass-loading model for TS location consistent with *Voyager 2* observations. ***MHD pressure from a 3D global heliospheric model that takes into account time-dependent SW effects over an 11-year solar cycle. ****MHD pressure that reflects a 50% increase in solar wind dynamic pressure, observed by *IBEX* to have arrived in the heliosheath by late 2016 (McComas et al. 2018). Finally, we note that Fuselier et al. (2012), Schwadron et al. (2011), Livadiotis et al. (2013), and Schwadron et al. (2014) were converted from LOS-integrated pressures using a nominal heliosheath thickness of ~30 AU.

Using a multiple-linear regression technique to infer all-energy LOS-integrated pressures from *IBEX*-observed ENA spectra, Livadiotis et al. (2013) estimated a partitioning of thermal (~6%), PUI (~63%), and energetic particle pressures (~31%) relative to the total pressure (~210 fPa) in the heliosheath. We adopt this scaling in order to estimate values for our non-observed quantities: PUI's and ACR's. Since it is the PUIs in the heliosheath that have been heated and ionized by the TS, we use a nominal value of ~30 AU for comparison to the LOS-integrated pressures.

Overall, our inferred pressures appear high compared to those of other data-driven models. There are several possible explanations for this. First, there could be other particle populations that contribute to the pressure which are currently unaccounted for in the simple partitioning shown in Table 1. For example, plasma electrons likely make some contribution to the pressure, although the extent of their possible contribution is unknown. Also, internal heating in the heliosheath could enhance the total pressure beyond that which is presently accounted for in most models. Or perhaps there is an extension to the suprathermal tail that is beyond *IBEX*'s measurement range. Another possibility is that the heliosheath thickness that we use to compare the LOS-integrated pressures is simply too large. Although we use a nominal thickness of 30 AU, perhaps the heliosheath could actually be thinner during this time because of movement of the termination shock. Second, the temperature in the VLISM is not directly measured – perhaps it is somewhat higher than the nominal value of 20,000 K that we have chosen. Even so, temperature only weakly contributes – at 40,000 K the total effective pressure that we predict is on the order of ~242 fPa. Third, we assume an adiabatic polytropic index to calculate pressure (Equation 4). Perhaps $\gamma$ is larger (possibly indicative of heating), or maybe the distribution is non-Maxwellian, as evidenced from *IBEX* measurements (Livadiotis et al. 2011).

Finally, we recall that our derivation of the GMIR's average speed assumes that it is the same in the two *Voyager* directions. Schwadron et al. (2014) found large-scale asymmetries in the heliosphere's global structure, resulting from a non-uniform pressure exerted by the interstellar magnetic field in its draping around the heliopause. They identified that the largest compression – and most enhanced pressure region – was located between the TS and HP at a location ~20° south and slightly offset to the port side from the interstellar upwind direction. Moreover, McComas and Schwadron (2014) suggested that this offset pressure maximum likely produced asymmetric plasma flows, providing a natural explanation of the unexpected flow directions in the heliosheath observed by *Voyager 2* (Richardson & Decker, 2014).

Nonetheless, it is unclear how much the GMIR's propagation time between *Voyager 2* and *Voyager 1* might be affected by latitudinal differences. The solar wind transitions from slow to fast at mid-latitudes during solar minimum (~400 to 700 km/s; McComas et al. 1998), and the PUI spectrum exhibits a latitudinal dependence (e.g., Fuselier et al. 2012). However, the two *Voyagers* are at opposing latitudes (roughly $\pm 35°$; Table 3) and Fuselier et al. (2012) found the low energy spectra (0.15 keV) to be nearly identical in the *Voyager 1* and *Voyager 2* directions. Schwadron et al. (2014)'s results also show that the LOS-integrated pixels in the two *Voyager* directions are similar. Furthermore, previous studies using 1-D and North-South spherically symmetric propagation models have proven successful in explaining *Voyager 2* and *Voyager 1* GMIR observations out to ~104 AU (see e.g., Webber et al. 2009). Therefore, we believe that latitudinal effects are likely small compared to other potential sources of uncertainty.

## 5. Discussion & Conclusion

Figure 3 presents daily averages of GCR, magnetic field and plasma data measured by *Voyager 2* and *Voyager 1*. The GCR disturbance in the heliosheath (Figure 3a) is accompanied by enhancements in the magnetic field (Figure 3b) and dynamic pressure (Figure 3c), signifying the passage of a GMIR (e.g., Burlaga et al. 2016; Richardson et al. 2017), but without indicators of a shock. Notably, the magnetic field drops well ahead of the dynamic pressure (~2012.7 vs. ~2012.8), and the dynamic pressure (~30 fPa) is at least a factor of nine smaller than our inferred total pressure (~270 fPa; Table 1). Although the specific details of these occurrences merit further investigation, they provide supplementary evidence that most of the energy is not in the plasma flow; the plasma dynamics inside the heliosheath are driven by populations other than the measured thermal solar wind core.

The contrast between the fields in the heliosheath (Figure 3b) and VLISM (Figure 3e) provides perhaps the best evidence of the differences between these two very different plasma regimes. For example, the heliosheath's magnetic fields are roughly four times weaker and marked by large fluctuations that are indicative of its strong turbulence (Burlaga et al. 2009). In contrast, the compressed fields of the VLISM are fairly smooth, reflecting its weak turbulence (Burlaga et al. 2015, 2018b).

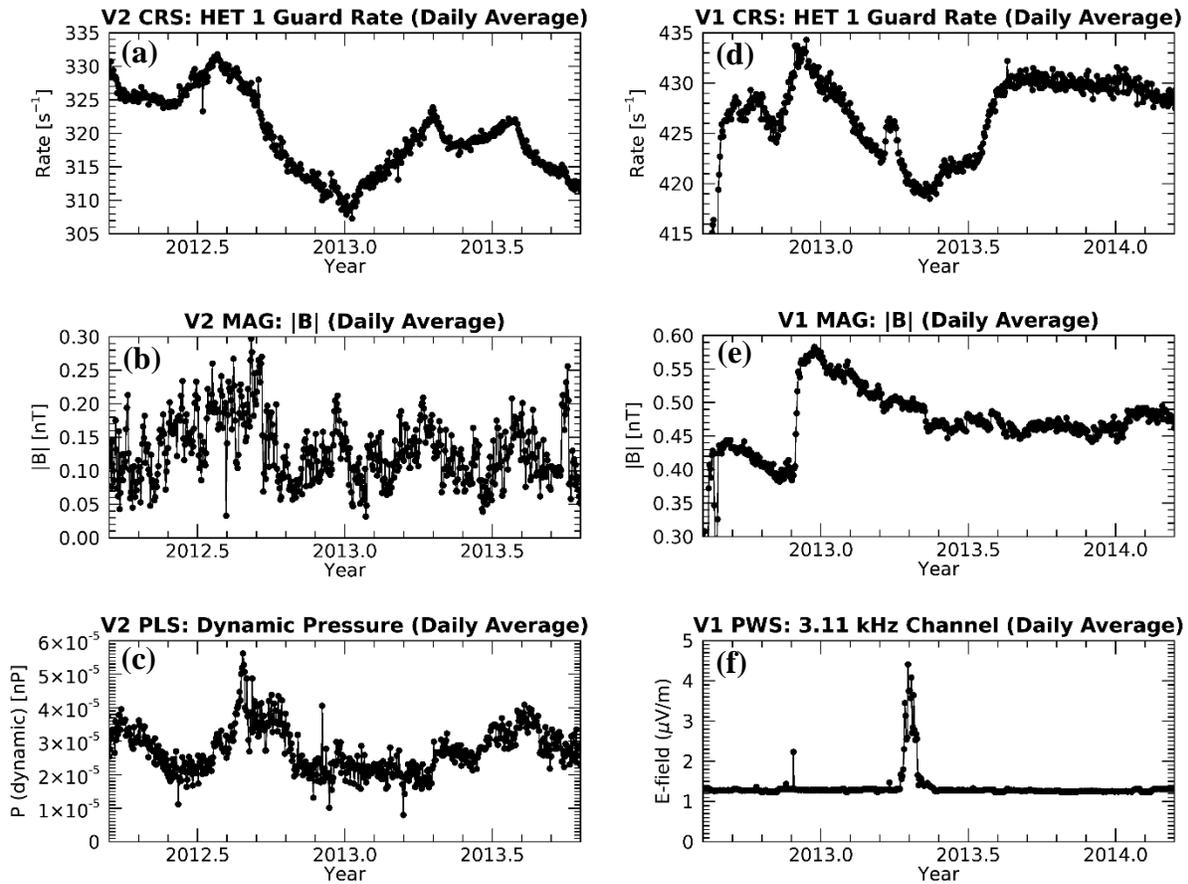

**Figure 3.** GCRs, magnetic fields, and plasma parameters in the heliosheath (a-c) and VLISM (d-f) measured by *Voyager 2* and *Voyager 1*, respectively. (a, d) ≳ 20 MeV proton-dominated GCRs from CRS omnidirectional rates. (b, e) Magnetic field strength observed by MAG (from publicly-available data: https://omniweb.gsfc.nasa.gov/coho/form/voyager1.htm; https://omniweb.gsfc.nasa.gov/coho/form/voyager2.htm). c) Solar wind dynamic pressure measured by PLS (*Voyager 2* only; from publicly-available data: http://web.mit.edu/space/www/voyager.html) and f) electric field response to plasma oscillations seen by PWS in their spectrum analyzer's 3.11 kHz channel (*Voyager 1* only; from publicly-available PLS data: http://www-pw.physics.uiowa.edu/voyager/).

Additionally, GCRs in the heliosheath may exhibit a weaker correlation with their local field (Figures 3a & 3b) than GCRs in the VLISM (Figures 3d & 3e), although this requires a more formal analysis. Such factors are important to understand, especially given that the particle response in the heliosheath is isotropic, while it is anisotropic in the VLISM (Krimigis et al. 2013; Rankin et al. 2019). In principle, the response of cosmic rays to a GMIR or pressure wave could result from the interplay of four competing processes: 1) trapping via reduced diffusion inside the GMIR, 2) acceleration by the shock's compression, 3) enhanced adiabatic cooling in the downstream shocked plasma, or 4) a gradient of the diffusion coefficient in front of the GMIR. Enhanced adiabatic cooling appears to be the dominant (though not the only) process for producing the GCR anisotropy in the VLISM (e.g., Kóta & Jokipii 2017; Rankin et al. 2019) while this is evidently not the case in the heliosheath. Nevertheless, understanding how well the first adiabatic invariant is preserved in the VLISM has important implications for a variety of physical processes (see, e.g. Burlaga & Ness, 2016; Zank et al. 2017; Schwadron & McComas 2017; Mostafavi & Zank 2018a, 2018b) and certainly merits further investigation with respect to the GCRs.

We also note that there is a clear particle enhancement (Figure 3d; ~2012.25) followed by a plasma oscillation (Figure 3f; ~2012.3) that occurs in the middle of *Voyager 1*'s GCR anisotropy event, which has no counterpart in the heliosheath. One explanation is that it might be caused by a remote shock that does not reach the spacecraft (e.g., Gurnett et al. 2015). Alternatively, Kóta & Jokipii (2017) successfully recreated both the anisotropy and these secondary enhancements via a shock geometry that deviated from the simple spherical case (see their Figures 4 & 5). Since the "shock spike" feature is not observed by *Voyager* 2, perhaps it reflects a distortion of the wave in its transmission across the across the HP arising from the non-uniform compression of the fields.

Although we have made some attempt to consider the effects of temporal variations in the heliosheath's pressure dynamics, there is still much work to be done. Clearly, a synthesis of *Voyager* and *IBEX* observations together are needed to significantly increase our understanding of this topic. *IBEX* regularly observes time variation in the morphology of the globally-distributed flux (see e.g., Schwadron et al. 2018 for the first 9 years of LOS-integrated pressure observations), and the *Voyagers* have a long history of measuring MIRs, GMIRs, and other transient disturbances in situ at their various locations through the heliosphere and beyond (e.g., McDonald et al. 1981; Burlaga et al. 1985; Gurnett et al. 1993; Whang & Burlaga 1995; Burlaga & Ness 1998; Paularena et al. 2001; Burlaga et al. 2003; Richardson et al. 2006; Webber et al. 2009; Burlaga et al. 2011; Luo et al. 2011). In addition to the 2012 events used in this study, *Voyager 2* observed several GMIRs and MIRs in the heliosheath from 2015 to 2018 (e.g., Burlaga et al. 2018a, 2019), and *Voyager 1* measured larger, more extended GCR anisotropy episodes starting in 2015 (e.g., Rankin et al. 2019) and 2018. Likewise, *IBEX* has now collected 10 years of data. These could serve as starting points to extend the current study.

Finally, we note that recent measurements by *Voyager* and *IBEX* reflect occurrences that differ significantly from previous observations. Over a relatively short time (~6 months) in late 2014, the solar wind dynamic pressure increased by 1 AU. By late 2017, *IBEX* observed enhanced ENA emissions indicative of the pressure front's arrival in the heliosheath (McComas et al. 2018). Simulations by McComas et al (2018) and Zirnstein et al. (2018) indicated that this abrupt change in pressure potentially moved the termination shock outward by ~7 AU and the heliopause by ~2 AU, and they expect that the expanding front will likely continue to be observed over the following many years. Moreover, accounting for time-propagation delays between ENA arrival at 1 AU and the expected arrival in the heliosheath, McComas et al. (2018) find their observations consistent with that of the large pressure pulse measured by *Voyager 2* in late-2015 – the largest enhancement in solar wind dynamic pressure so far measured in the heliosheath (Richardson et al. 2017). Meanwhile, in the VLISM, *Voyager 1* observed an unusual, "distinct large-scale magnetic feature" marked by a ~35-day increase in the field (starting on DOY 346 of 2016), followed by a ~665-day decay, which Burlaga et al. (2019) suggest is probably related to the pressure front observed by *IBEX*. These findings, in conjunction with an unusually long, shallowly-sloped, GCR anisotropy that began in 2018 and continues up to present in the VLISM certainly necessitates a careful follow-on analysis.

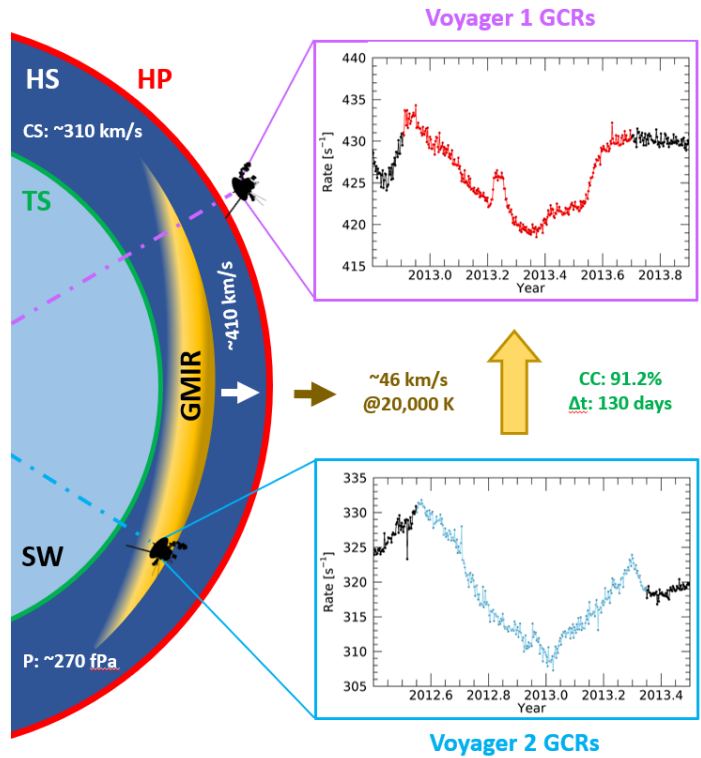

**Figure 4.** Schematic diagram depicting a spherical GMIR that generates a GCR intensity disturbance at *Voyager 2* and transmits a pressure pulse across the HP, producing a causally-related disturbance at *Voyager 1*. For $T_{VLISM} = 20,000$ K, we find that the GMIR's speed (white arrow and text) during this time period is ~410 km/s, the transmitted wave's speed (brown arrow and text) is ~46 km/s and the heliosheath's sound speed (CS) and pressure (P) are ~310 km/s and ~270 fPa, respectively (white text).

Our findings are summarized in the schematic diagram shown in Figure 4. Observations support that a GMIR at likely passed through the heliosheath and collided with the HP, thereby transmitting a pressure pulse into the VLISM. The timing of the resulting GCR disturbances, the locations of the spacecraft, and the combination of local particle, field, and plasma observations enabled us to infer a range of sound speeds and total effective pressures in the heliosheath as a function of temperature in the VLISM. Topics that we suggest for further investigation include: 1) utilizing similarities and differences in the heliosheath and VLISM to better understand the GCR anisotropy, 2) extending this work to future events, including the 2015 GMIR/anisotropy and the large-scale pressure wave event of 2017, 3) utilizing multi-spacecraft observations of heliospheric transients to better understand possible spatial variations (including latitudinal and longitudinal dependencies as well as other possible non-radial effects), and 4) engaging in a multi-year, multi-spacecraft study of time-variability in the heliosheath's dynamics.

Thus, for the first time, we have used measurements of a *Voyager 2* to *Voyager 1* transient to derive sound speed and total effective pressure in the heliosheath and have considered these findings in light of data-driven models, including *IBEX* ENA observations. We look forward to continuing our evaluations of both *Voyager*'s in-situ observations and *IBEX*'s global measurements – ideally in combination with multiple past, present, and future spacecraft – as together they can generate greater insight about the heliosphere and its interaction with the interstellar medium.

## Acknowledgements

We thank E. Zirnstein and P. Mostafavi for helpful discussions about PUI pressures and densities, shocks, and wave speeds in the heliosheath. We are grateful to both the *Voyager* and *IBEX* teams. This work was supported by NASA Grant 80NSSC18K0237 and by the *IBEX* mission, which is part of NASA's Explorer Program. John D. Richardson was supported under NASA contract 959203 from the Jet Propulsion Laboratory to the Massachusetts Institute of Technology.

# Appendix A: Cross-Correlation Analysis

The cross-correlation function measures the correlation between two data sets, $a(t)$ and $b(t)$, displaced by variable $t$. For discrete sets of data points:

$$(a \star b) = \frac{1}{k_{max} - k_{min}} \sum_{k=k_{min}}^{k=k_{max}} \frac{(a[k] - \mu_a)(b[k+t] - \mu_b)}{\sigma_a \sigma_b} \quad (5)$$

where $\sigma_a$ and $\sigma_b$ are the standard deviations of $a_k$ and $b_k$, respectively. The function achieves its maximum value, $CC = (a \star b)_{max}$, for some lag, $\Delta t$, within a reasonable time window: $t \in [t_{min}, t_{max}]$. In this study, $a[k]$ represents CRS's HET 1 omnidirectional guard rate on *Voyager 1* ($\gtrsim$ 20 MeV; proton-dominated), and $b[k]$ represents the equivalent rate on *Voyager 2*. Additionally, $\Delta t$ represents the time difference between the GCR responses to the pressure wave at *Voyager 2*, followed by *Voyager 1*, and CC represents the degree to which the events at the two spacecraft are correlated.

By stepping selected *Voyager 1* data through a larger time-window of the *Voyager 2* observations ($t2_{window} \in [2010.0, 2013.7]$), we accumulate sets of $(a \star b)$. Since the onset time of the GCR disturbance at *Voyager 1* is not obvious, we additionally perform calculations for a range of possible start times, within the window: $t1_{start} \in [2012 \text{ DOY } 290, 2013 \text{ DOY } 39]$. Consequently, our CC ultimately reflects the global maximum of all $(a \star b)$. Interestingly, we find that the optimum *Voyager 1* start time occurred on DOY 331 of 2012 – very near to the arrival of the shock observed by MAG (roughly DOY 335; Burlaga et al. 2013). This is physically consistent with the idea that most VLISM anisotropies form via particle trapping and energy loss in the adiabatically-expanding fields of the shocked plasma downstream (Kóta & Jokipii 2017; Rankin et al. 2019).

To determine the statistical uncertainties in CC and $\tau$, we randomly draw samples from Poisson distributions, $a \sim P(\mu_a, N_a)$ and $b \sim P(\mu_b, N_b)$, to produce different random sets of $\overline{a}, \overline{b}$. We then perform $(\overline{a} \star \overline{b})$ to find the maximum correlation, $\overline{CC} = (\overline{a} \star \overline{b})_{max}$ and corresponding lag time, $\overline{\Delta t}$. Finally, after a million iterations, we generate distributions of $\overline{CC}$'s and $\overline{\Delta t}$'s from which we calculate the standard deviations, $\sigma_{ab}$ and $\sigma_\tau$, that describe our uncertainties. We report our final results in the form: $CC \pm \sigma_{ab}$ and $\Delta t \pm \sigma_{\Delta t}$. In the end, we arrive at: $CC = 91.2\% \pm 2.1\%$ and $\Delta t = 130 \pm 5$ days using time windows of $t_{V1} \in [2012 \text{ DOY } 331, 2013 \text{ DOY } 255]$ and $t_{V2} \in [2012 \text{ DOY } 201, 2013 \text{ DOY } 125]$ for *Voyager 1* & *Voyager 2*, respectively.

## Appendix B: Results

| $T_{VLISM}$ [K] | $v_{VLISM}$ [km/s] | $v_{HS}$ [km/s] | $cs_{HS}$ [km/s] | $P_{total}$ [fPa] |
|---|---|---|---|---|
| 7,500 | 42.1 ± 4.0 | 421 ± 43 | 329 ± 34 | 292 ± 60 |
| 8,000 | 42.3 ± 4.0 | 421 ± 43 | 328 ± 34 | 291 ± 60 |
| 8,500 | 42.4 ± 4.0 | 420 ± 43 | 327 ± 34 | 290 ± 60 |
| 9,000 | 42.6 ± 4.0 | 419 ± 43 | 327 ± 34 | 289 ± 59 |
| 9,500 | 42.7 ± 4.1 | 419 ± 43 | 326 ± 34 | 287 ± 59 |
| 10,000 | 42.9 ± 4.1 | 418 ± 43 | 325 ± 33 | 286 ± 59 |
| 10,500 | 43.1 ± 4.1 | 418 ± 43 | 325 ± 33 | 285 ± 59 |
| 11,000 | 43.2 ± 4.1 | 417 ± 43 | 324 ± 33 | 284 ± 59 |
| 11,500 | 43.4 ± 4.1 | 416 ± 43 | 324 ± 33 | 283 ± 58 |
| 12,000 | 43.5 ± 4.1 | 416 ± 43 | 323 ± 33 | 282 ± 58 |
| 12,500 | 43.7 ± 4.1 | 415 ± 43 | 322 ± 33 | 281 ± 58 |
| 13,000 | 43.9 ± 4.2 | 415 ± 42 | 322 ± 33 | 280 ± 58 |
| 13,500 | 44.0 ± 4.2 | 414 ± 42 | 321 ± 33 | 279 ± 58 |
| 14,000 | 44.2 ± 4.2 | 414 ± 42 | 321 ± 33 | 278 ± 57 |
| 14,500 | 44.3 ± 4.2 | 413 ± 42 | 320 ± 33 | 277 ± 57 |
| 15,000 | 44.5 ± 4.2 | 412 ± 42 | 320 ± 33 | 276 ± 57 |
| 15,500 | 44.6 ± 4.2 | 412 ± 42 | 319 ± 33 | 275 ± 57 |
| 16,000 | 44.8 ± 4.3 | 411 ± 42 | 318 ± 33 | 274 ± 57 |
| 16,500 | 44.9 ± 4.3 | 411 ± 42 | 318 ± 33 | 273 ± 56 |
| 17,000 | 45.1 ± 4.3 | 410 ± 42 | 317 ± 33 | 272 ± 56 |
| 17,500 | 45.2 ± 4.3 | 410 ± 42 | 317 ± 33 | 272 ± 56 |
| 18,000 | 45.4 ± 4.3 | 409 ± 42 | 316 ± 33 | 271 ± 56 |
| 18,500 | 45.6 ± 4.3 | 409 ± 42 | 316 ± 33 | 270 ± 56 |
| 19,000 | 45.7 ± 4.3 | 408 ± 42 | 315 ± 32 | 269 ± 55 |
| 19,500 | 45.9 ± 4.4 | 408 ± 42 | 315 ± 32 | 268 ± 55 |
| 20,000 | 46.0 ± 4.4 | 407 ± 42 | 314 ± 32 | 267 ± 55 |
| 20,500 | 46.2 ± 4.4 | 407 ± 42 | 314 ± 32 | 267 ± 55 |
| 21,000 | 46.3 ± 4.4 | 406 ± 42 | 313 ± 32 | 266 ± 55 |
| 21,500 | 46.4 ± 4.4 | 406 ± 42 | 313 ± 32 | 265 ± 55 |
| 22,000 | 46.6 ± 4.4 | 406 ± 42 | 313 ± 32 | 264 ± 54 |
| 22,500 | 46.7 ± 4.4 | 405 ± 41 | 312 ± 32 | 263 ± 54 |
| 23,000 | 46.9 ± 4.5 | 405 ± 41 | 312 ± 32 | 263 ± 54 |
| 23,500 | 47.0 ± 4.5 | 404 ± 41 | 311 ± 32 | 262 ± 54 |
| 24,000 | 47.2 ± 4.5 | 404 ± 41 | 311 ± 32 | 261 ± 54 |
| 24,500 | 47.3 ± 4.5 | 403 ± 41 | 310 ± 32 | 260 ± 54 |
| 25,000 | 47.5 ± 4.5 | 403 ± 41 | 310 ± 32 | 260 ± 54 |
| 25,500 | 47.6 ± 4.5 | 403 ± 41 | 309 ± 32 | 259 ± 53 |
| 26,000 | 47.8 ± 4.5 | 402 ± 41 | 309 ± 32 | 258 ± 53 |
| 26,500 | 47.9 ± 4.5 | 402 ± 41 | 309 ± 32 | 258 ± 53 |
| 27,000 | 48.1 ± 4.6 | 401 ± 41 | 308 ± 32 | 257 ± 53 |
| 27,500 | 48.2 ± 4.6 | 401 ± 41 | 308 ± 32 | 256 ± 53 |
| 28,000 | 48.3 ± 4.6 | 401 ± 41 | 307 ± 32 | 256 ± 53 |
| 28,500 | 48.5 ± 4.6 | 400 ± 41 | 307 ± 32 | 255 ± 53 |
| 29,000 | 48.6 ± 4.6 | 400 ± 41 | 307 ± 32 | 254 ± 52 |
| 29,500 | 48.8 ± 4.6 | 399 ± 41 | 306 ± 32 | 254 ± 52 |
| 30,000 | 48.9 ± 4.6 | 399 ± 41 | 306 ± 31 | 253 ± 52 |
| 30,500 | 49.0 ± 4.7 | 399 ± 41 | 305 ± 31 | 252 ± 52 |
| 31,000 | 49.2 ± 4.7 | 398 ± 41 | 305 ± 31 | 252 ± 52 |
| 31,500 | 49.3 ± 4.7 | 398 ± 41 | 305 ± 31 | 251 ± 52 |
| 32,000 | 49.5 ± 4.7 | 397 ± 41 | 304 ± 31 | 250 ± 52 |
| 32,500 | 49.6 ± 4.7 | 397 ± 41 | 304 ± 31 | 250 ± 51 |
| 33,000 | 49.7 ± 4.7 | 397 ± 41 | 304 ± 31 | 249 ± 51 |
| 33,500 | 49.9 ± 4.7 | 396 ± 41 | 303 ± 31 | 249 ± 51 |
| 34,000 | 50.0 ± 4.7 | 396 ± 41 | 303 ± 31 | 248 ± 51 |

| 34,500 | 50.2 ± 4.8 | 396 ± 41 | 303 ± 31 | 248 ± 51 |
| 35,000 | 50.3 ± 4.8 | 395 ± 40 | 302 ± 31 | 247 ± 51 |
| 35,500 | 50.4 ± 4.8 | 395 ± 40 | 302 ± 31 | 246 ± 51 |
| 36,000 | 50.6 ± 4.8 | 395 ± 40 | 301 ± 31 | 246 ± 51 |
| 36,500 | 50.7 ± 4.8 | 394 ± 40 | 301 ± 31 | 245 ± 51 |
| 37,000 | 50.8 ± 4.8 | 394 ± 40 | 301 ± 31 | 245 ± 50 |
| 37,500 | 51.0 ± 4.8 | 394 ± 40 | 300 ± 31 | 244 ± 50 |
| 38,000 | 51.1 ± 4.9 | 393 ± 40 | 300 ± 31 | 244 ± 50 |
| 38,500 | 51.2 ± 4.9 | 393 ± 40 | 300 ± 31 | 243 ± 50 |
| 39,000 | 51.4 ± 4.9 | 393 ± 40 | 299 ± 31 | 243 ± 50 |
| 39,500 | 51.5 ± 4.9 | 392 ± 40 | 299 ± 31 | 242 ± 50 |
| 40,000 | 51.6 ± 4.9 | 392 ± 40 | 299 ± 31 | 242 ± 50 |

**Table 2.** Listing of results described in the text and plotted in Figure 2. Derived quantities include the transmitted wave's speed in the VLISM, $v_{VLISM}$, and the GMIR's speed, $v_{HS}$, the sound speed, $cs_{HS}$, and the total pressures, $P_{total}$, in the heliosheath, all as a function of VLISM temperature, assuming negligible VLISM radial bulk flow.

|  | **Event** | **Location** | **Year** | **DOY** | **Latitude** | **Longitude** | **Radial Distance** | **Spacecraft Speed** |
|---|---|---|---|---|---|---|---|---|
| **Voyager 1** | HP Crossing | Heliopause | 2012 | 238 | +35.0° | 255.0° | 121.58 AU | 17.0 km/s |
| **Voyager 2** | Onset of GCR Disturbance | Heliosheath | 2012 | 201 | -34.2° | 289.6° | 99.05 AU | 14.9 km/s |
| **Voyager 1** | Onset of GCR Disturbance | VLISM | 2012 | 331 | +35.0° | 255.1° | 122.49 AU | 17.0 km/s |

**Table 3.** Summary of the location and times of: 1) *Voyager 1*'s HP crossing, 2) the onset of the GCR disturbance at *Voyager 2*, and 3) the onset of the GCR disturbance at *Voyager 1*. The locations are listed in solar ecliptic coordinates.

| $\|B\|_{ave}$* (HS) [nT] | $n_{ave}$ (HS) [cm$^{-3}$] | $T_{ave}$ (HS) [K] | $v(r)_{ave}$ (HS) [km/s] | $P_{thermal}$ (HS) [fPa] | $P_{dynamic}$ (HS) [fPa] |
|---|---|---|---|---|---|
| 0.116 ± 0.001 | 21.6 × 10$^{-4}$ ± 0.2 × 10$^{-4}$ | 52,490 ± 314 | 88.3 ± 0.3 | 3.12 ± 0.03 | 28.1 ± 0.2 |

**Table 4.** *Voyager 2* PLS Plasma parameters used to derive the sound speed in the heliosheath (HS). In order to calculate the GMIR's speed over the distance it covers in the heliosheath, we use data averaged from the time of the onset of the event (2012, DOY 201), up to roughly 11 days before *Voyager 2*'s own HP crossing on 2018, DOY 309 (Brown et al. 2018). These values are obtained using publicly-available PLS (http://web.mit.edu/space/www/voyager.html) and MAG (https://omniweb.gsfc.nasa.gov/coho/form/voyager1.html) data. We report statistical uncertainties for all heliosheath measurements, without taking instrumental uncertainties into account. As such, our uncertainties for the sound speed and total effective pressure are of the order of ±10% and ±20%, respectively. If we incorporate MAG's reported uncertainty of ±0.03 nT for |B|, this drives our uncertainties to roughly ±20% for the sound speed and greater than ±50% for the total effective pressure. Nevertheless, we find the reported uncertainty in |B| to be much larger than the data implies and are unsure of its origin, as it appears to have been the same value since the *Voyager* mission began. *The magnetic field data is only available through 2015-365.